\documentclass[prb,aps,superscriptaddress,floatfix,tightenlines,showpacs,twocolumn]{revtex4}

\usepackage{graphicx}
\usepackage{dcolumn}   
\usepackage{bm}        
\usepackage{amssymb}   
\usepackage{amsmath}
\usepackage{url,CJK}
\usepackage{layout,color}
\usepackage{epsfig}
\usepackage{graphicx}
\usepackage{epstopdf}
\usepackage{booktabs}
\usepackage{float,braket}
\usepackage[hyperindex,breaklinks]{hyperref}

\def\be{\begin{equation}}
\def\ee{\end{equation}}
\def\ba{\begin{eqnarray}}
\def\ea{\end{eqnarray}}

\begin{document}
\begin{CJK}{UTF8}{gbsn}
\title{Construction of Wannier Functions in Disordered Systems}

\author{Junbo Zhu(竺俊博)}
\affiliation{International Center for Quantum Materials, School of Physics, Peking University, Beijing 100871, China}
\author{Zhu Chen(陈竹)}
\affiliation{Institute of Applied Physics and Computational Mathematics, Beijing 100088, China}
    \author{Biao Wu(吴飙)}   \email{wubiao@pku.edu.cn}
      \affiliation{International Center for Quantum Materials, School of Physics, Peking University, Beijing 100871, China}
      \affiliation{Collaborative Innovation Center of Quantum Matter, Beijing 100871, China}
      \affiliation{Wilczek Quantum Center, College of Science, Zhejiang University of Technology, Hangzhou 310014, China}

\begin{abstract}
We propose a general method of constructing Wannier functions in disordered systems directly out of energy eigenstates. This method consists of two successive operations: (i) a phase transformation setting the proper localization center; (ii) the mixing of adjacent states in energy to sufficiently minimize the spread of the Wannier functions. The latter operation can be well approximated by a band matrix, further facilitating the calculation. Detailed implementation of our method is illustrated with one dimensional systems; the generalization to higher dimensions is straightforward.
\end{abstract}
\date{\today}
\maketitle
\end{CJK}

\section{Introduction}
\label{sec:introduction}
In periodic systems, e.g., crystalline, the Bloch eigenstates, which are extended in space while localized in energy, naturally form a complete basis. A proper unitary transformation of the Bloch basis can be made to give rise to an equivalent representation with the basis of Wannier functions (WFs), which are localized in space \cite{Wannier37}. The WFs offer an insightful picture of chemical bonds, serve as the center of modern theory of polarization \cite{Resta94,Vanderbilt93}, and are the basis for an efficient linear-scaling algorithms in electric-structure calculations \cite{Goedecker99,Galli96}. In particular, they act as bedrock in the construction of model Hamiltonian, such as the Bose-Hubbard model in cold atomic physics \cite{Waghmare95,Anderson2001}.

It has been proven that for periodic system with time reversal symmetry, there always exists a set of exponentially localized WFs, which is associated with the analytical behavior of the corresponding Bloch states \cite{Kohn59,Cloizeaux64_1,Cloizeaux64_2,Nenciu83,Nenciu93,Nenciu98,Marzari07}. Nonetheless, the arbitrariness in the phase as well as the weights of the Bloch states makes the unitary transformation non-unique, thus prohibiting the achievement of the `optimal' WFs. A variety of methods have been proposed to remove such arbitrariness, among which Kohn's scheme \cite{Kohn59} is exact however applicable only in one dimensional system with inversion symmetry. The widely-used maximal localization procedure of Marzari and Vanderbilt relies on a numerical minimization \cite{Marzari97,Mostofi08}.

Recently, efforts have been made to generalize the Wannier representation to disordered system \cite{Kivelson82,Kohn73,Kohn93,Ceperley10,Marzari12,Uehlinger13}.  The motivation is of two folds. First, there are always disorders in real materials, making the understanding of disordered system being of fundamental importance. Second, most recently, the development in cold atom experiments \cite{Bloch2008,White09} offers the possibility of a full control of disorder, both on the type and the intensity, calling for the accurate mapping from continuous system to lattice model. In the presence of point defect or a more general weak disorder \cite{Kohn73,Kivelson82,Kohn93}, studies support the existence of a set of localized WFs. Further research reveals that the essential obstruction of the construction of WFs is of topological origin and suggests that the localization property of WFs is a direct consequence of the existence of an energy gap rather than the perfect crystalline order \cite{Nenciu98,Nenciu93,Kivelson82}. This implies the existence of localized basis for a general disordered system as long as the energy gap is preserved. Such localized basis is called generalized Wannier functions (GWFs).

Unlike the periodic case, how to obtain the optimal GWFs is still open though the existence has been proven. One solution arises by treating the disordered system a unit cell embedded in a larger periodic system \cite{Marzari12} and expecting the bulk properties being captured through such approximation. Following this line, the Wannier function as well as the existing methods of construction still applies. Other attempts focus on the direct dealing  with the disordered systems themselves. In a recent work \cite{Ceperley10}, a method based on the projection operator \cite{Anisimov07} is proposed by Zhou and Ceperley to deal with general disordered optical lattice.

In this work we propose an alternative method to construct GWFs of non-periodic system. We deal with directly the eigenstates of the disordered system and decompose the unitary transformation into two successive operations: (i) a phase transformation; (ii) a transformation relating the states adjacent in energy space. The latter is approximated by a band matrix with the elements determined through a minimization of the spread of GWFs. The implementation of our method is illustrated in detail with
one dimensional systems, where exponentially localized GWFs are achieved.
Our method has potential applications in a wide range of areas. In the field of ultracold atomic gas,
it facilitates the mapping of continuous systems to discrete models \cite{Bloch2008,Uehlinger13,Chen14}. In
the field of condensed matter physics,  our method can be used to calculate GWFs
for disordered systems and materials with large unit cells.

This paper is organized as follows. In Sec. \ref{sec:method}, we describe the general scheme 
of our method. In Sec. \ref{sec:examples},  we give a
detailed illustration on the implementation of our method in the one dimensional systems.   Afterward, we discuss at what strength of the disorder our method begins to fail in Sec. \ref{sec:discussion}. We finally conclude in Sec. \ref{sec:conclusion}.

\section{Method}
\label{sec:method}
Wannier functions form a complete orthonormal basis of a Hilbert space; they are related to  the energy eigenstates $\phi_m(\mathbf{r})$ via a unitary transformation,
\begin{equation}\label{eq1}
W_n(\mathbf{r})=\sum_{m}S_{nm}\phi_{m}(\mathbf{r})\,,
\end{equation}
where $m$ labels different energy eigenstates and $n$ marks different WFs.
Such a unitary transformation also holds  within a subspace of the Hilbert space which
is spanned by a group of energy eigenstates that are well separated from other
energy states by finite energy gap(s).
In the familiar case of periodic system, the energy eigenstates
are Bloch waves. When only a single isolated Bloch band is considered,
the WFs take the form of the Fourier transformation of Bloch states,
\begin{equation}\label{eq:2}
W_n(\mathbf{r})=\frac{1}{\sqrt{N}}\sum_{\mathbf{k}_m} e^{-i\mathbf{k}_m\cdot\mathbf{R}_n}\psi_{\mathbf{k}_m}(\mathbf{r})
\end{equation}
Comparing Eq. (\ref{eq:2}) with Eq. (\ref{eq1}), we find $S_{nm}$ being simply the phase term $\frac{1}{\sqrt{N}} e^{-i\mathbf{k}_m\cdot\mathbf{R}_n}$ and energy eigenstate $\phi_m$ being Bloch states $\psi_{\mathbf{k}_m}$. $\mathbf{k}_m$'s are the Bloch wave vectors and are discrete due to the periodic box boundary condition. $R_n$'s  are the lattice locations.

We consider a slightly perturbed periodic  system described by $H=H_0+\lambda H_1$
($\lambda \ll 1$) with $H_0$ periodic and $H_1$  disordered.  This situation was studied by Geller and Kohn \cite{Kohn93} with the perturbation theory. It is suggested that up to the first order of $\lambda$, the transformation matrix remains the same as its periodic counterpart while the formal Bloch states are substituted by the so called Generalized Bloch Functions (GBFs), i.e., $|\psi_\mathbf{k}\rangle=|\psi_\mathbf{k}^0\rangle+\lambda\sum_{\mathbf{k}\neq\mathbf{k'}}\frac{\langle \psi_{\mathbf{k'}}^0|H_1|\psi_{\mathbf{k}}^0 \rangle}{E_{\mathbf{k}}^0-E_{\mathbf{k}'}^0} $. Clearly, the main correction arises from those states which are close in energy due to the factor $\frac{1}{E_{\mathbf{k}}^0-E_{\mathbf{k}'}^0}$. This perturbation theory does not hold for cases with stronger disorder. However, the idea that transformation from periodic system to disordered system is supposed to be continuous sheds light on how to deal with
systems with strong disorder.

We propose that the transformation $S$ in Eq. (\ref{eq1}) can be decomposed into two
successive operations as
\begin{equation}\label{eq:S_decompose}
S_{nm}=\sum_{l}T_{nl}U_{lm}
\end{equation}
where $T_{nl}=\frac{1}{\sqrt{N}}e^{-i\mathbf{k}_l\cdot{\mathbf{R}_n}}$ is a phase term, being  the same as that in Eq. (\ref{eq:2}).  Though physically speaking, a `wave vector' $\mathbf{k}_l$ no longer exists in the presence of disorder, such a phase transform $T$ is proven to be helpful to set the Wannier centers correctly, paving the way for further processing as stated in the following section.  $U$ is a general matrix mixing different $\phi_m$'s. Inspired by the perturbational treatment of Geller and Kohn \cite{Kohn93}, we assume that $U$ is a band matrix, since only states sufficiently close in energy contribute the most. This greatly facilitates our numerical calculation. Finally, the explicit form of $U$ is obtained under the requirement that GWFs are sufficiently localized. This is accomplished by the  minimization of a Wannier spread function $\Omega$. In this work a common form of $\Omega$ is used\cite{Marzari97},
\begin{equation}\label{eq:Omega}
\Omega=\sum_n[\langle W_n|r^2|W_n\rangle-\langle W_n|r|W_n\rangle^2]=\overline{r^2}-\bar{r}^2
\end{equation}
Actually other definitions are tried such as  $\Omega=\sum_n[\int|W_n(r-R_n)|^4dr]$ and they work as well as Eq. (\ref{eq:Omega}).

We emphasize that our method proceeds directly with energy eigenstates and shall apply to systems with strong disorder, which is out of the reach of perturbational theory. Furthermore, an explicit minimization is used and supposed to provide better localization than that from imaginary evolution \cite{Ceperley10}.

\begin{figure}[ht]
\includegraphics[width=\linewidth]{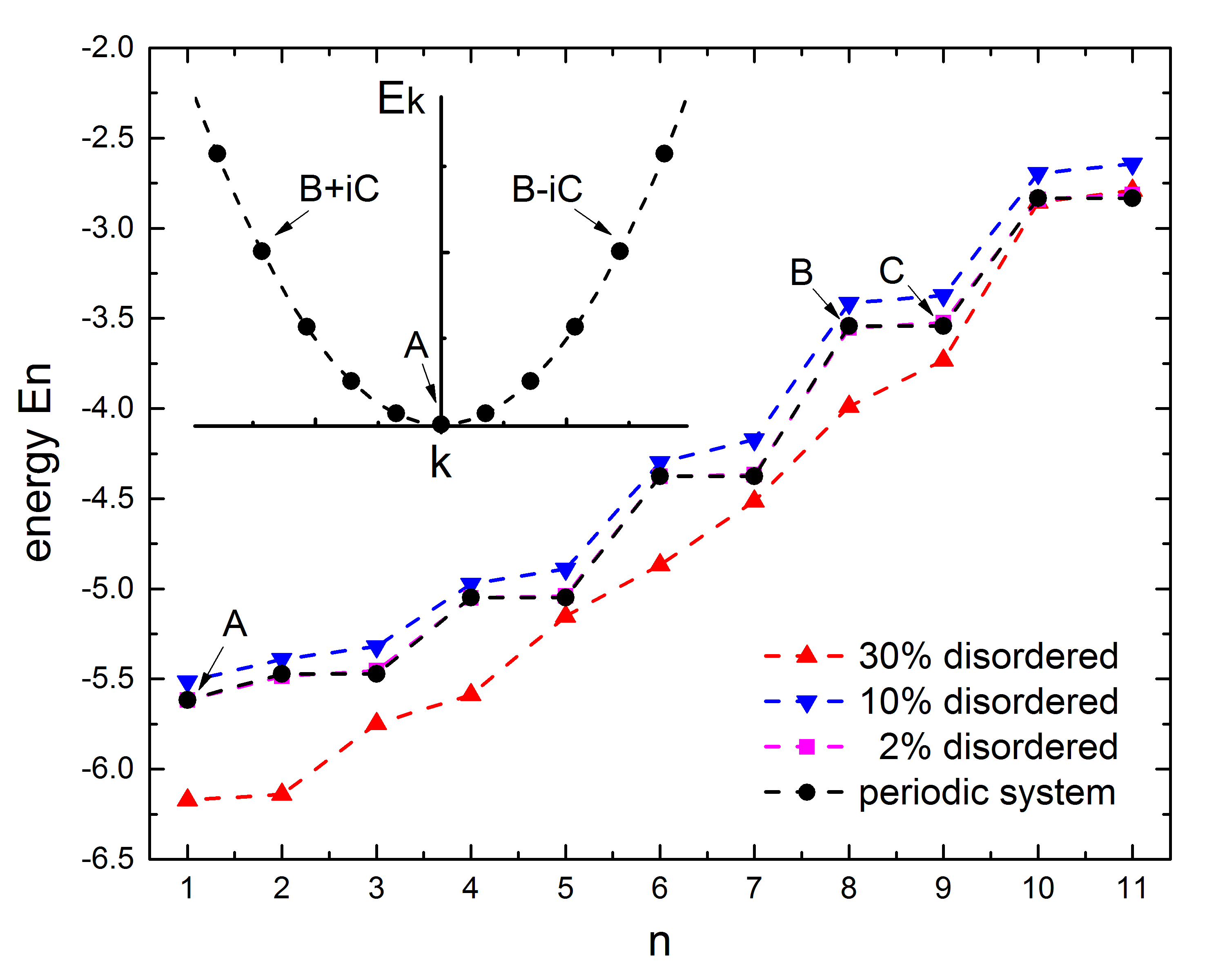}
\caption{\label{fig:energy} (Color online) Energies of eigenstates calculated in real space by finite-difference method for 11 wells (cosine shape) with periodic boundary condition. The black dots correspond to periodic case, while the magenta squares, blue down triangles and red up triangles correspond to potentials with 2\%, 10\% and 30\% disorder in well depth. The inset is a schematic diagram shows the energy plot in k space for periodic system. Points A, B and C show typical correspondence between eigenstates calculated in real space and Bloch states in the $k$ space.}
\end{figure}

\section{Examples}
\label{sec:examples}

\begin{figure*}[ht]
\includegraphics[width=0.49\linewidth]{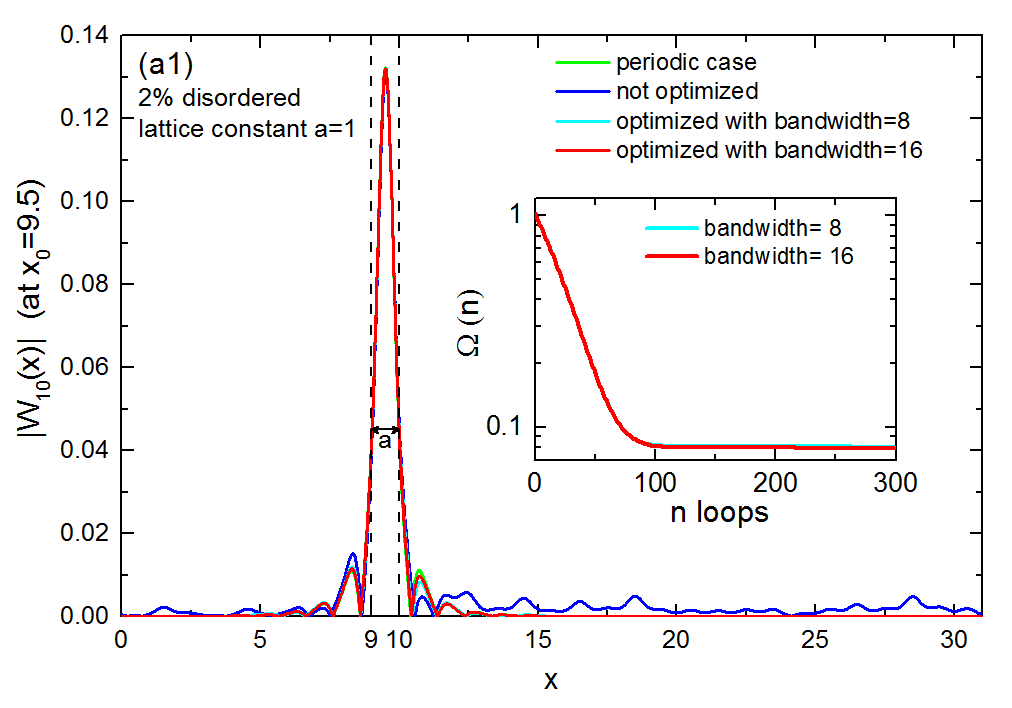}\includegraphics[width=0.49\linewidth]{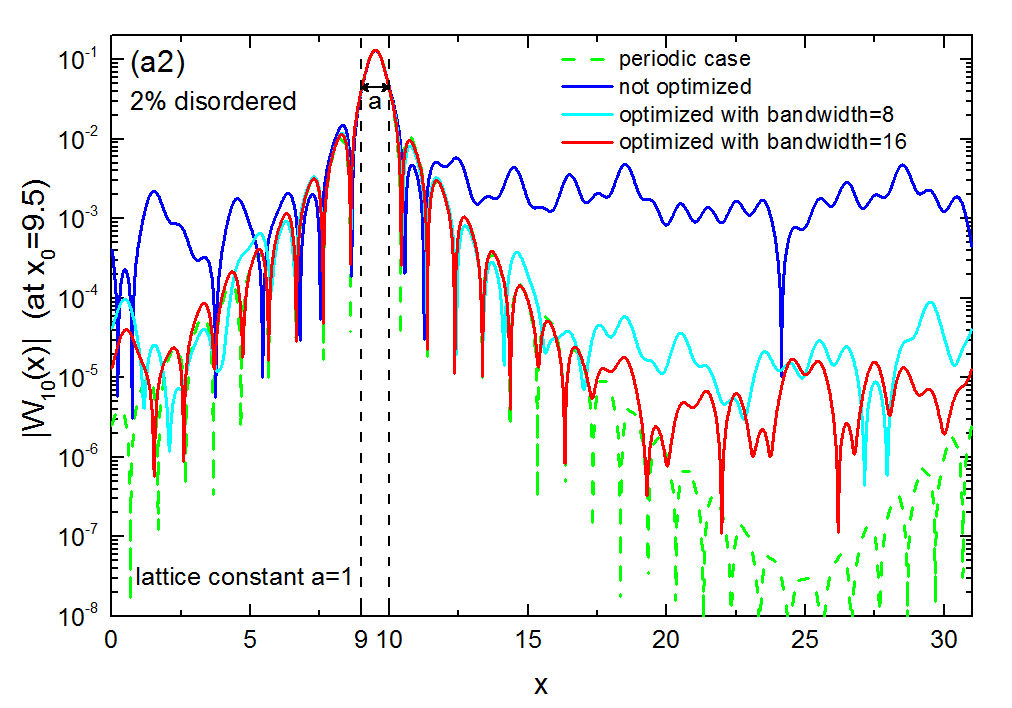}\\
\includegraphics[width=0.49\linewidth]{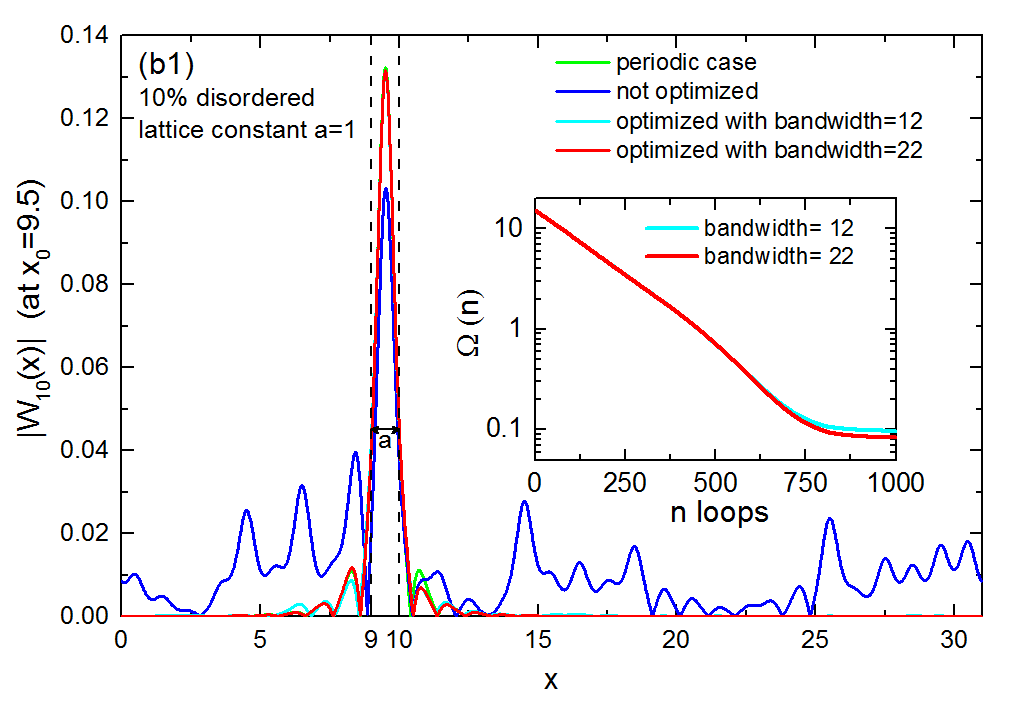}\includegraphics[width=0.49\linewidth]{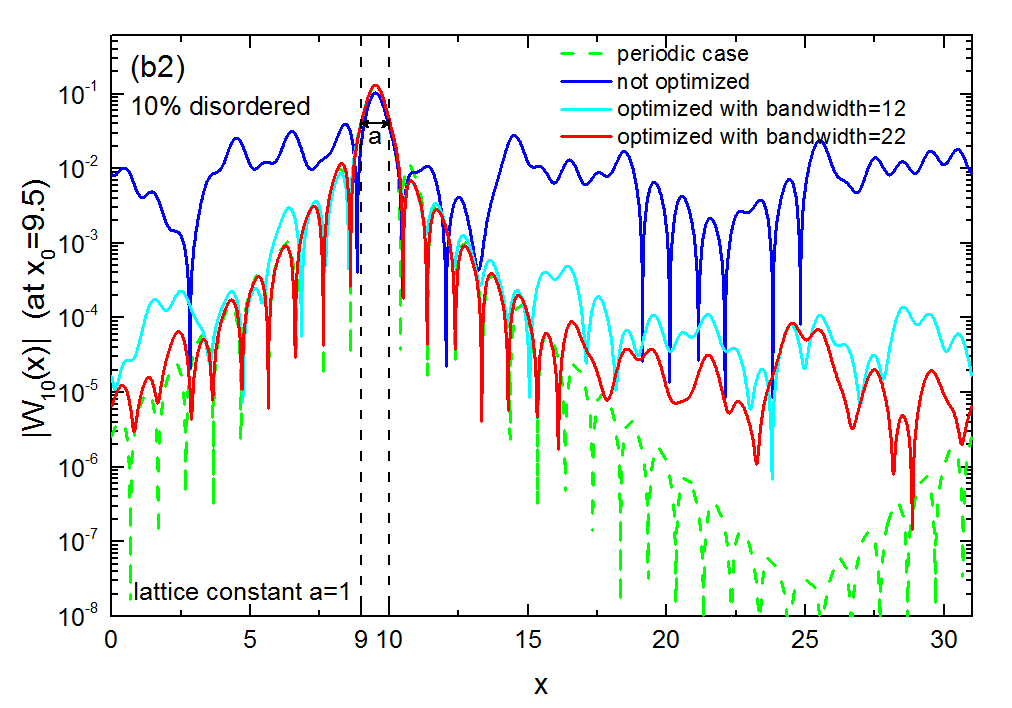}
\caption{\label{fig:result}(Color online) Wannier functions in disordered potentials in Eq.(\ref{potential}).
$a=1$ and $A=5$. (a)$2\%$ disordered system ($\eta=0.02$). (b)$10\%$ disordered system ($\eta=0.1$). (a1) Absolute value of WFs located at $x_0=9.5$ (the dashed straight lines denote the unit cell WF located). The green line is the WF of the periodic system, plotted as an ideal case for comparison and other lines are of disordered system; the blue line is the WF after initial $TU_0$ transformation but before functional optimization; the cyan line is optimized WF using bandwidth b=8 band matrix; and the red line is also optimized WF but using bandwidth b=16 band matrix. The inset shows log plot of the localization functional $\Omega$ as a function of iteration loop number n and 300 is the total number of optimization loops; again the cyan line for b=8 and the red line for b=16.(a2) shows log plot of the same WF in (a1). (b1)(b2) show plots for 10\% disordered system corresponding to (a1)(a2), but the cyan line is for b=12, the red line for b=22 and total optimization loop number is 1000.}
\end{figure*}

In this section we  illustrate  our method with examples. Without loss of generality, we choose the
disordered potential as a series of  cosine-type wells of random depths,
\begin{equation} \label{potential}
V_n(x)=A_n[\cos(2\pi x/a)-1]\,,
\end{equation}
where $n$ marks the index of wells and $a$ is the lattice constant. The disorder manifests itself in the lattice strength,
\begin{equation}\label{eq:amplitude}
A_n=A[1+\eta\cdot\mathfrak{R}_n]\,,
\end{equation}
where $A$ is a constant for well depth and  $\mathfrak{R}_n$ denotes a sequence of random numbers between -0.5 and 0.5. $\eta$ denotes the relative strength of disorder. For example, we refer to $\eta=0.1$ as a 10\% disorder.  We use $A=5$ and $a=1$ in our examples.

\begin{figure*}[ht]
\includegraphics[width=0.49\linewidth]{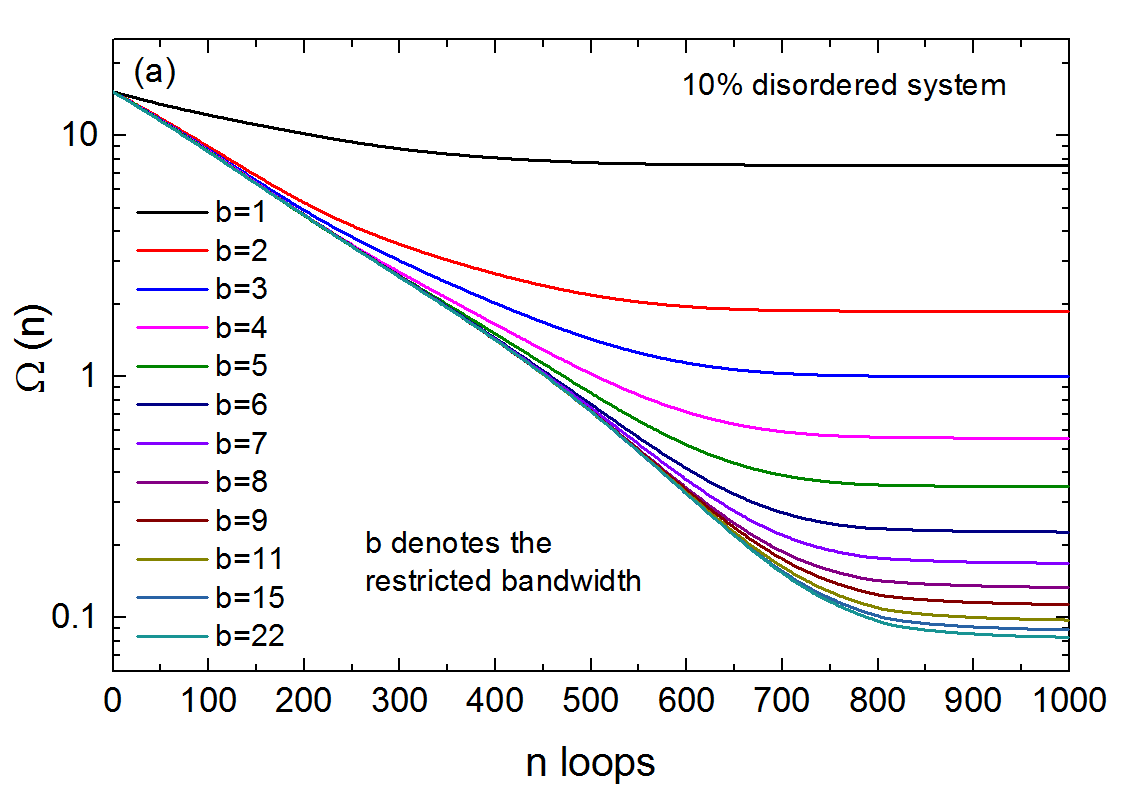}
\includegraphics[width=0.49\linewidth]{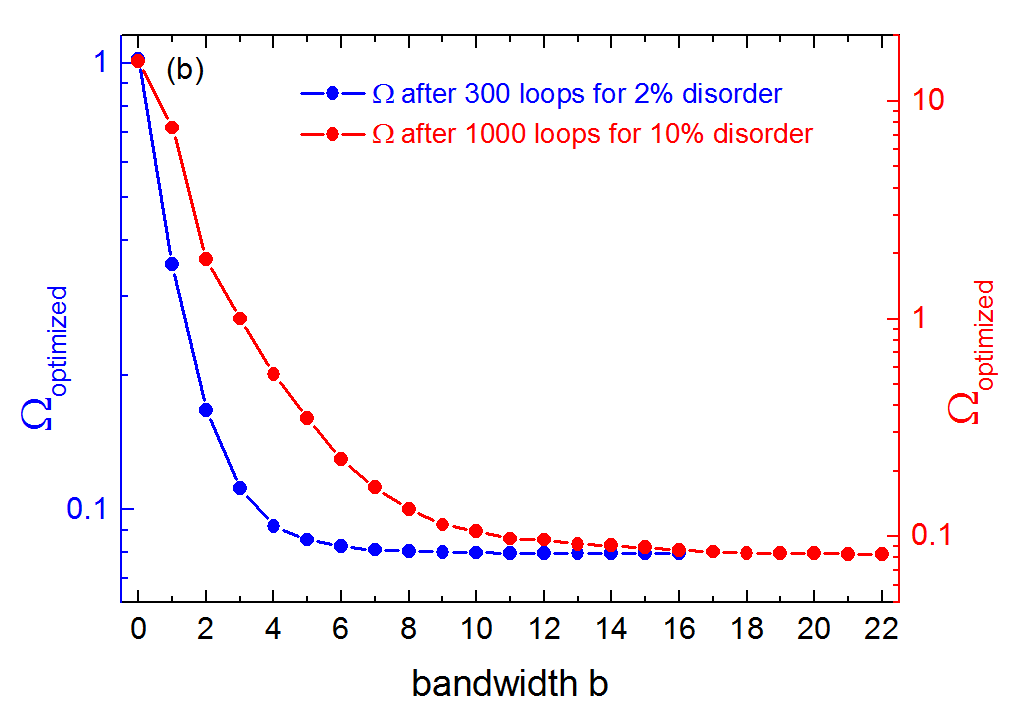}
\caption{\label{fig:band}(Color online) (a) Log plot of the localization functional $\Omega$ as a function of iteration loop number n for 10\% disordered system. $\Omega$ all decreases exponentially when loop number n increases and larger bandwidth leads to better optimized value of $\Omega$. (b) Log plot of optimized $\Omega$ as a function of bandwith b. The blue line is for the 2\% disordered case  
corresponding to the $y$ axis on the left hand side and the red line is for the 10\% disordered 
case corresponding to the $y$ axis on the right hand side.}
\end{figure*}

We begin with the periodic system, i.e., $\eta=0$.  The system is solved directly
in real space for 11 wells with periodic boundary condition. The obtained energy levels are ordered in increasing sequence. The group of the lowest eigen-energies are shown in Fig. \ref{fig:energy}
and they constitute a sampling of the complete lowest Bloch band with each point corresponding to $E_{\mathbf{k}}$ with different $k$. The same energy spectrum is shown in the inset
in the $k$ space as black dots. The degenerate pairs (e.g., B,C ) in the spectrum is related by a time reversal operation as $E_{\mathbf{k}}=E_{-\mathbf{k}}$. We emphasize that the eigenstates obtained in
our calculations are real functions and the degenerate pair are the real and imaginary part of Bloch functions ($\psi_{\mathbf{k}}, \psi_{\mathbf{-k}}$) rather than Bloch functions themselves. A typical example is given in the inset, where $B$ and $C$ mark the corresponding eigenstates respectively and their linear combinations leads to Bloch functions. Such fact makes the transformation $U$ block diagonal as $U_0=\textrm{diag}(1,U_0^1,U_0^2\cdots)$ with each block
\begin{equation}\label{eq:U0}
\quad U^j_0=\left(\begin{array}{cc}1&i\\1&-i\end{array}\right)
\end{equation}
such that $\psi_{k_i}=\sum_j\{U_0\}_{ij} \phi_j$. Then Eq. (\ref{eq1}) becomes
\begin{equation}\label{eq:3periodic}
W_n(\mathbf{r})=\sum_{m,j}T_{nm}\{U_0\}_{mj}\phi_{j}(\mathbf{r})
\end{equation}
where $\quad T_{nm}=\frac{1}{\sqrt{N}}e^{-i\mathbf{k}_m\cdot\mathbf{R}_n}$. Note that an additional procedure following the scheme of Kohn \cite{Kohn59} could further eliminate the sign uncertainty of eigenstates, thus completes our method without any further calculation. We refer to Appendix \ref{sec:initial} for more details.

When $\eta\neq0$, disorder manifests itself in the breaking of degeneracy between pairs
such as $B$ and $C$, which is clearly seen from the 2\%, 10\% and 30\% disordered cases in Fig. \ref{fig:energy}.
The tendency that stronger disorder leads to a severer modification of energy spectrum 
is also seen.
Since we only aim to provide a simple and clear illustration of our proposal, disorders of moderate strength with the energy sequence as well as a finite band gap being preserved are of concern in this section. Further discussion on strong disorder is given in the following section. In the present case, the transformation $T$ is fixed as $T_{nm}=\frac{1}{\sqrt{N}}e^{-ik_m\cdot R_n}$ with $k_m$ still being the wave vector of the periodic counterpart. $U$ has to be determined numerically with the bandwidth $b$ a variational parameter itself (e.g., $b=1$ for a tridiagonal matrix). We adopt a steepest descent algorithm to iteratively achieve the minimization of $\Omega$ with $U=U_0$ initially. More details are presented in Appendix \ref{sec:optimization}.

Typical generalized Wannier functions localized in specific wells are obtained.
As an example, the GWF at  localization center $x_0=9.5$ is plotted in Fig. \ref{fig:result}.
Comparison is made between periodic ($\eta=0$) and weak disordered ($\eta=0.02$) cases.
Fig. \ref{fig:result}(a2) is the log plot of  Fig. \ref{fig:result}(a1) for a clearer sight. We find
that the GWFs (blue and red lines) are well localized within the well (see the dashed lines),
sharing the same localization center as well as a similar exponentially decaying behavior
as their periodic counterpart (see the green lines). Apparent deviations appear
in tails where the localization seem to be polynomial. However, the distant tails reside almost 10 wells away from the localization center, leaving the subsequent overlap integral negligibly small and are expected not to cause significant effect. In this sense, we conclude that the GWFs show a comparable localization property as the WFs in accordance with the anticipation of perturbation analysis \cite{Kohn59}. The GWFs obtained through an initial $TU_0$ transformation but without further functional minimization procedure are also plotted (see the blue lines). Though poor localization is shown, the proper Wannier centers are captured, based on which further optimization turns out to be possible.

Results of stronger disorder ($\eta=0.1$) are presented in Fig. \ref{fig:result}(b1,b2). On the one hand, the corresponding GWFs show an exponential localization comparable to the periodic counterpart as well. On the other hand, stronger disorder manifests itself in the severer deviation of the distant tail as well as the appearance of comparable peaks around the Wannier center (see blue lines) before optimization. Following this trend, larger disorder is expected to cause the smearing of the Wannier center and the prohibition of the further optimization. This  will be discussed later in more detail.

The Wannier spread function $\Omega$s as function of iteration number $n$ are plotted in the insets of in Fig. \ref{fig:result}(a1,b1) for both disordered cases. Rapid convergence with an exponential form and a final achievement to saturations for all choices of bandwidths is clearly seen. This provides a complementary evidence to the validity of our proposal. Note the trend that larger bandwidth leads to smaller saturation of $\Omega$ (thus more localized GWFs) with increasing iterations is resulted from larger functional space for optimization other than the specific algorithm we choose. In other words, the optimal choice of GWFs has been obtained within a given bandwidth.

We emphasize that although only GWFs localized in a specific cell ($x_0=9.5$) are shown and compared for simplification, we calculate all the GWFs at the same time. Optimized GWFs for disordered case, despite the extent of disorder, all exhibit the same level of localization property as that for the periodic case within about 15 nearby unit cells. This shows the validity of  our method.

In addition, the band matrix approximation of $U$ is also checked. Recall the intuitive picture drawn from the perturbation theory that the eigenstates close in energy  contribute most to the construction of GBFs. Two facts are thus expected for a general disorder. First, a matrix of finite bandwidth is sufficient for the achievement of optimal WFs. Second, the bandwidth increases as the increment of disorder strength. The former fact is verified in Fig. \ref{fig:band}(a), where the evolution of Wannier spread $\Omega$ with respect to iterations are shown for different bandwidths. On the one hand, more localized GWFs which can be inferred from decreasing $\Omega$ are achieved as bandwidth $b$ increases. This is consistent with physical intuition. On the other hand, the approaching to saturations of $\Omega$ with respect to increasing $b$ indicates the final convergence. This property is further demonstrated from Fig. \ref{fig:band}(b)  where the spread $\Omega$ as a function of bandwidth is shown for weak and relative stronger disorder. In weak disorder case (blue dots), the existence of an optimal bandwidth $b=8$ can be inferred which leads to a satisfactory description of GWFs. In strong disorder case (red dots), the optimal bandwidth turn out to be $b=12$. This provides a direct evidence for the second fact. If we look back at Fig. \ref{fig:result} (a2,b2), the almost overlapping of results from different bandwidths (red and cyan lines) supports the choice of optimal bandwidth. The bandwidth restriction can be further released as a trade-off between localization and calculation.

\section{Discussion}
\label{sec:discussion}
Our method is rooted in a new perspective that the unitary transformation $S$ from energy eigenstates
to Wannier functions in periodic systems consists of two operations $T$ and $U_0$: $S=TU_0$.
To construct GWFs in disordered systems is to generalize $U_0$ to a band matrix $U$.

The initial $TU_0$ transformation sets the correct localization centers for the GWFs, then
the GWFs are made more localized by a straightforward optimization. It appears  that it is the $TU_0$ transformation that underlies the wide adaptability of our method. With this in mind, we turn to stronger disordered cases to further inspect our method.

\begin{figure}[tp]
\includegraphics[width=\linewidth]{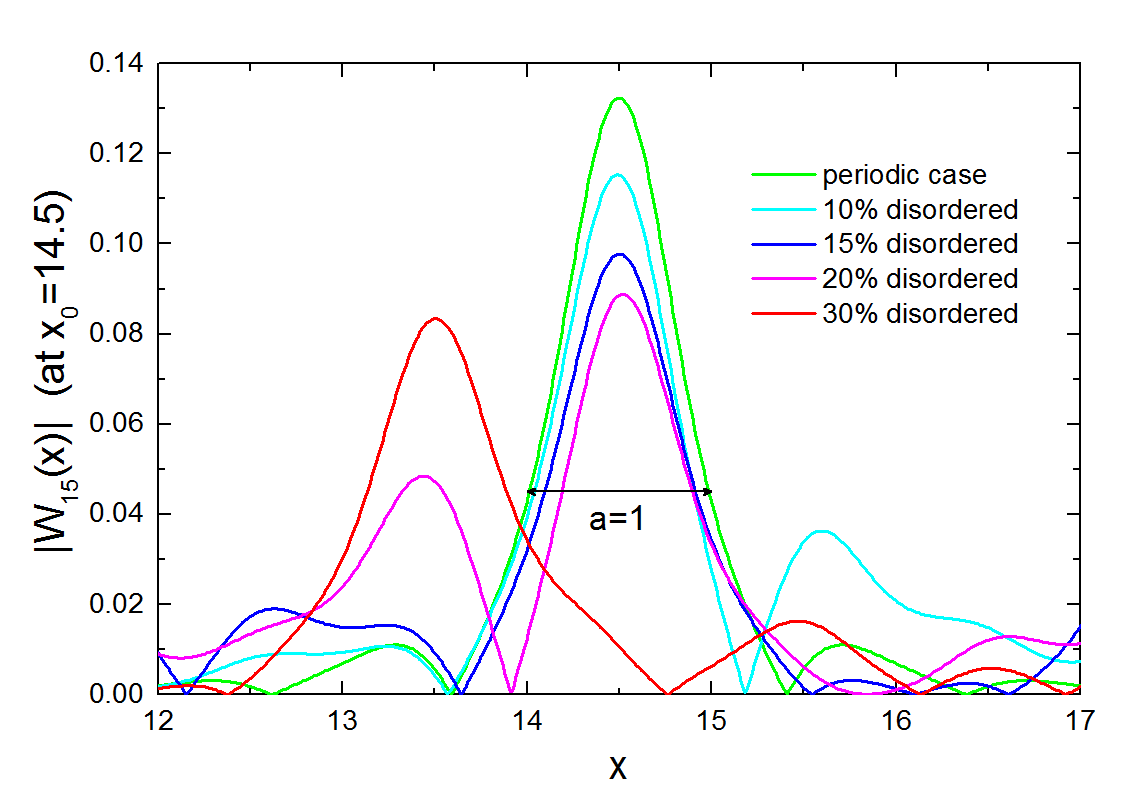}
\caption{\label{fig:limit} (Color online) WFs after initial transformation for different extents of disordered systems. All of them are expected to localized at x0=14.5 where the corresponding atom located (lattice constant a=1). Clearly when disorder is as large as 30\% or larger, this initial WF loses the localization center and thus fails to work as a good starting point of functional optimization.}
\end{figure}

We focus on functions just after the initial $TU_0$ transformation, without being further optimized, for a specific cell located at $x_0=14.5$. A series of such GWFs corresponding to different strengths of disorder are illustrated in Fig. \ref{fig:limit} for comparison. Apparently, the $TU_0$ transformation captures the anticipated localization center for cases with disorder from 0 up to 20\%. However, if the adjacent cell at $x_0=13.5$ is also inspected, clues may be found that an additional peak arises with increasing strength in the wake of increasing disorder. This additional peak becomes comparable with that in the localization center, e.g., the 20\% case (see the pink line). Eventually, in a case with 30\% disorder, the additional peak prevails and the anticipated localization is completely shifted to $x_0=13.5$. In this case, the following optimization procedure fails to achieve an exponentially localized GWFs.

We conjecture that this phenomenon results from the band mixing that some wells may be sufficiently shallow with its lowest energy level residing in a higher band of the system, leading to the insufficient description with only the lowest band. So the failure in the very example is nothing but an indication that higher bands should be taken into account.

Note that in the examples shown above the average well depth is 10, i.e. constant $A$ in Eq. (\ref{eq:amplitude}) equals to 5, and our method works well for disorder up to about 15\%. While numerical calculations also shows that for smaller average well depth, our method can endure larger disorder, e.g. for systems with average well depths equal to 6 and 2, it works well for disorder as large as 25\% and 40\% respectively. Moreover, although periodic boundary condition is adopted in the examples, our method also works well with infinite boundary condition, which is often the case in cold atom experiment.

One may doubt that whether the Anderson localization (AL) \cite{Anderson58} which is inevitable in sufficiently disordered system (in 1D, it appears as long as the presence of disorder) will prevent us from getting a `Bloch-like' function from the $TU_0$ transformation on the eigenstates. Our numerical results show that within disorder strength up to 10\%, eigenstates well expand through more than 100 unit cells, indicating the appearance of AL in a region greatly larger than the WFs, which are supposed to be localized in only several unit cells. Thus with regard to the disorder considered in the present text, no significant effect caused by AL should be taken into account.

\section{Conclusion}
\label{sec:conclusion}
We have proposed an alternative method to construct a set of exponentially localized general Wannier functions in disordered systems. Detailed illustration is given in terms of examples, showing the advantages of high efficiency in calculation as well as broad applicability from weakly perturbed system to system with large disorder beyond perturbation method. The band matrix approximation provides a balance between calculation and accuracy. The generalization to higher dimension is straightforward, though only 1D is of consideration for simplicity. Our proposal can find its applications in a wide range of areas, providing a new method of constructing GWFs for disordered systems and large supercells in real materials and also facilitating the mapping from continuous to discrete model in cold atomic physics.

\acknowledgments 
This work is supported by the National Basic Research Program of China (Grants No. 2013CB921903 and No. 2012CB921300) and the National Natural Science Foundation of China (Grants No. 11274024, No. 11334001, and No. 11429402).

\begin{appendix}
\section{Method Details}
In this Appendix, details of the eliminating of phase arbitrariness in Bloch functions as well as the steepest descend method which have been omitted in the main text are represented. This finally completes our proposal.
\subsection{Initial $TU_0$ Transformation}
\label{sec:initial}
In Sec. \ref{sec:examples}, Eq. (\ref{eq:U0}) is used to transform eigenstates calculated in real space to Bloch states. However, freedom remains during this process described as:
\begin{equation}
\left(\begin{array}{c}\psi_{k^1_i}\\ \psi_{k^2_i}\end{array}\right)=U_0^i\left(\begin{array}{c}\phi_{2i}\\ \phi_{2i+1}\end{array}\right)
\end{equation}
with $k^1_i=-k^2_i=\pm k_i$. But the choice of $U_0^i$ still remains some freedoms. Apart from an overall arbitrary phase, $U_0^i$ could be either $A$ or $AB$, given that
\begin{equation}
A=\left(\begin{array}{cc}1&i\\1&-i\end{array}\right);B=\left(\begin{array}{cc}0&1\\1&0\end{array}\right)
\end{equation}
resulting in different choices of $\psi_{+k_i}$ or $\psi_{-k_i}$. Thus additional procedure is needed to obtain a proper sequence of $\psi_k$ and eliminate the corresponding arbitrary overall phase. The procedures can be described as follows:

\emph{Step 1: Make $\phi_{m}(x=0)$ is positive (real itself).}\par
\emph{Step 2: Transform from $\phi$ to $\psi_{k}$ in pairs. (choose either A or AB to be $U_0^i$)}\par
\emph{Step 3: Check the sign of wave vector k from $\psi_{k}$ and resort the sequence of $\psi_{k}$ to be consistent with T matrix in Eq. (\ref{eq:S_decompose}).}\par
\emph{Step 4: Gauge choice: make $\psi_{k}(x=x0)$ is real and positive. (x0 is the lowest point of V)}\par

The gauge in step 4 follows the suggestion of Kohn \cite{Kohn59,Kohn73} in constructing WFs for periodic and symmetric potential, which still works in our scheme. After these four steps, for periodic system, $S=TU_0$ directly transforms eigenstates to WFs without further calculation, while for non-periodic systems, we take $S=TU_0$ serve as an initial transformation, followed by a numerical minimization of the spread function.

Note that in step 3, wave vector $k$ is determined as the phase change from $\psi(x)|_{x_1}$ to $\psi(x)|_{x_1+a}$, where $a$ is the lattice constant.
For non-periodic system, similar `wave vectors' can be obtained by taking the average of `$k$ value' (obtain from every $x$) over the whole space. Such wave vector are used only in the present process with the the signs help to determine the propagation direction of this `Bloch-like' state. Note that $k_m$ used in $T_{nm}=e^{-ik_m\cdot R_n}$ is still the k value of the corresponding periodic case, which evenly divides the Brillouin zone.

\subsection{Functional Optimization}
\label{sec:optimization}
During the functional optimization of $\Omega$, successive operations on $U$ is exerted as
\begin{equation}
U_0\rightarrow U_1\rightarrow \cdots\rightarrow U_n,\quad
S=T\lim_{n\rightarrow\infty}U_n
\end{equation}

Algorithm as simple as Steepest Descent method (SD) turns out to work well in our examples. More efficient algorithms may be chosen, however, this is more of a technique problem and out the scope of the present paper.

During an update from $U_j$ to $U_{j+1}$, our task is to determine the steepest gradient direction of $\Omega$ as a function of $U$. We use
\begin{equation}
U_{j+1}=e^{dU_j}U_j
\end{equation}
where $dU_j$ is anti-Hermitian, i.e. $dU_{j}^{\dagger}=-dU_j$, which guarantees the unitarity  of $U_{j+1}$:
\begin{equation}
U_{j+1}U_{j+1}^\dagger=e^{dU_j}U_jU_j^\dagger(e^{dU_j})^\dagger=e^{dU_j}e^{-dU_j}=I
\end{equation}

As an anti-Hermitian matrix, only the elements in the upper band (or lower band) of $dU$ are  independent parameters. Usually $dU$ is set to be a band matrix with bandwidth $b$ (both upper and lower bandwidth). Then $e^{dU}$ and subsequent $U_{j+1}$ are in general matrix with bandwidth larger than $b$. However, if we keep to only the first order approximation, $b$ still make sense.
\end{appendix}

\end{document}